\documentclass[12pt,twoside]{article}
\usepackage{amsthm,amsmath,amssymb,amscd}
\DeclareMathOperator{\Rea}{Re}
\DeclareMathOperator{\Ima}{Im}
\DeclareMathOperator{\rank}{rank}
\DeclareMathOperator{\Symp}{Sp}
\DeclareMathOperator{\M}{M}

\DeclareMathOperator{\Hilb}{Hilb}
\DeclareMathOperator{\ch}{ch}
\DeclareMathOperator{\td}{td}
\numberwithin{equation}{section}

 \renewcommand{\tilde}{\widetilde}
 \newtheorem{theorem}{Theorem}[section]
 \newtheorem{df}[theorem]{Definition}
 
 \newtheorem{cor}[theorem]{Corollary}

 \newcommand{\tharg}[2]{\genfrac{[}{]}{0pt}{}{#1}{#2}}
 \newcommand{\abs}[1]{\lvert#1\rvert} 
\font\sectionfont=cmbx12 at 12pt
\font\myaddressfont=cmti12

\def\section#1{\vskip 1.5truepc\centerline{\hbox {{\sectionfont #1}}}
\vskip 1truepc\noindent\stepcounter{section}}
\font\myaddressfont=cmti12
\font\authorfont=cmti12 at 14pt
\font\titlefont=cmbx12 at 16pt

\begin{document}

\def\sec#1{\vskip 1.5truepc\centerline{\hbox {{\sectionfont
#1}}}\vskip1truepc\noindent\noindent\stepcounter{section}}

\def\newsubsec#1#2{\vskip .3pc plus1pt minus 1pt\noindent {\bf #1}
{\rm #2}}

\newcommand{\CC}{\mathbb{C}}
\newcommand{\RR}{\mathbb{R}}
\newcommand{\QQ}{\mathbb{Q}}
\newcommand{\ZZ}{\mathbb{Z}}
\newcommand{\HH}{\mathfrak{H}}
\newcommand{\NN}{\mathbb{N}}
\newcommand{\PP}{\mathbb{P}}
\newcommand{\frS}{\mathfrak{S}}

 \catcode`\@=11
\font\twelvemsb=msbm10 scaled 1200
\font\tenmsb=msbm10
\font\ninemsb=msbm7 scaled 1200%msbm9
\newfam\msbfam
\textfont\msbfam=\twelvemsb  \scriptfont\msbfam=\tenmsb
  \scriptscriptfont\msbfam=\ninemsb
\def\msb@{\hexnumber@\msbfam}
\def\Bbb{\relax\ifmmode\let\next\Bbb@\else
 \def\next{\errmessage{Use \string\Bbb\space only in math
mode}}\fi\next}
\def\Bbb@#1{{\Bbb@@{#1}}}
\def\Bbb@@#1{\fam\msbfam#1}
\catcode`\@=12

\renewcommand{\thefootnote}{\fnsymbol{footnote}}

\centerline{\titlefont K3 Surfaces, Igusa Cusp Form and String Theory}
\vskip 2pc
\centerline{\authorfont Toshiya Kawai}

\date{}

\vskip 3pc

\centerline{\sectionfont Abstract} 

\vskip 8pt 
\rm It has recently become apparent that the elliptic genera of K3
surfaces (and their symmetric products) are intimately related to the
Igusa cusp form of weight ten.  In this contribution, I survey this
connection with an  emphasis on string theoretic viewpoints.

% submitted to the proceedings of the Taniguchi symposium, 
% ``Topological Field  Theory, Primitive Forms and Related Topics'', 
% December 9th -- 13th, 1996

\section{1. Introduction}

This is a survey of somewhat mysterious connections found recently
between $K3$ surfaces and the Igusa cusp form of weight ten
$\chi_{10}$.  

Gritsenko and Nikulin [GN1] following the method of Borcherds [Bor1]
proved a remarkable infinite product formula of $\chi_{10}$.  The
exponents of the factors appearing in their infinite product coincide
with the coefficients in  the expansion of the elliptic genus of K3
surfaces.  The aim of this survey is to try to look at this
coincidence more carefully.

Borcherds-type infinite products appear naturally in some sort of
modular integrals associated with one-loop calculations of string
theory as pointed out by Harvey and Moore [HM]. They used the method
developed by Dixon, Kaplunovsky and Louis (DKL) [DKL].  In sect.6, I
present a similar calculation [Kaw1] for the case of our present
interest. The approach taken is again that of DKL, which may now
fairly be said to be standard, but I will be more careful about the
so-called ``degenerate orbits'' explaining the relevance of the
Kronecker limit formulas.

The Igusa cusp form $\chi_{10}$ [Igu] made its appearance in the past
in the two-loop calculations of critical bosonic string theory [Two].
Thus one is naturally led to the expectation that K3 surfaces might be
related to bosonic string. In fact, Vafa and Witten [VW] pointed out
the connection between K3 surfaces and bosonic string in the context
of G{\" o}ttsche's formula [Goe].  They gave a physical derivation of
the orbifold version [HH] of G{\" o}ttsche's formula. This method of
Vafa and Witten has recently been extended by Dijkgraaf {\it et al.}
[DMVV] for the computation of the generating function of the orbifold
elliptic genera of symmetric products of K3 surfaces.  In sect.7, I
give some interesting interpretation of their result taking the
relation between $\chi_{10}$ and the two-loop vacuum amplitude of
bosonic string seriously.

The two subjects treated in sect.6 and sect.7 are not logically
connected and it is desirable to gain a conceptual as well as rigorous
understanding of why they exhibit a parallelism.  However, there have
been interesting proposal and argument regarding this in conjunction
to string duality [DVV][DMVV].

We also note that in addition to the works mentioned, the Igusa cusp
forms have been used in other aspects of string theory [MS] [Kaw2]
[CCL].

Sects.2--5 are devoted to preliminary materials. In sect.2, we
review the definitions of  Siegel modular forms and Jacobi
forms. In sect.3, we summarize the fundamental properties of elliptic
genera. In sect.4, we investigate in detail the elliptic genus of K3
surfaces. In sect.5 we review the infinite product formula of
Borcherds, Gritsenko and Nikulin.

Finally, in sect.8, we point out what remains to be clarified.

\bigskip
\noindent{\bf Notation.}
\smallskip

\indent $\RR$: the set of real numbers.\\
\indent $\CC$: the set of complex numbers.\\
\indent $\ZZ$: the set of integers.\\
\indent $\NN$: the set of positive integers.\\
\indent $\NN_0$: the set of non-negative integers. \\
\indent$\M(n,\RR)$: the set of $n\times n$ real matrices.\\
\indent $\M(n,\CC)$: the set of $n\times n$ complex matrices.\\
\indent$\M(n,\ZZ)$: the set of $n\times n$ integer matrices.\\
\indent$I_n$: the $n\times n$ identity matrix.\\
\indent${\bf e}[x]$: $\exp(2\pi\sqrt{-1}x)$.\\

\section{2. Siegel modular forms and Jacobi forms}
First I recall some fundamental materials from the theory of Siegel
modular forms [Fre] and that of Jacobi forms [EZ].

\pagestyle{myheadings}
\markboth{{}\myaddressfont \hfil T. Kawai}
{\myaddressfont K3 surfaces, Igusa cusp form and string theory\hfil {} }

The Siegel upper space of genus $g \in \NN$ is defined by
\begin{equation}
  \HH_g :=\{ \Omega\in \M(g,\CC)\mid \Omega={}^t\Omega,\quad
  \Ima(\Omega) > 0\}\,.
\end{equation}
The symplectic group
\begin{equation}
  \Symp(2g,\RR):=\{\gamma\in \M(2g,\RR)\mid {}^t \gamma J \gamma=J\},\quad
  J= \left(\begin{array}{cc} 0 & I_g \\ -I_g & 0\end{array}\right)\,,
\end{equation}
acts on $\HH_g$ by
\begin{equation}
       \gamma \cdot \Omega= (A\Omega+B)(C\Omega+D)^{-1}\,,
\end{equation}
where 
\begin{equation}
  \gamma=\left(\begin{array}{cc} A&B \\ C&D \end{array}\right)\in
  \Symp(2g,\RR) ,\quad A,B,C,D \in \M(g,\RR)\,.
\label{gamma}
\end{equation}
This action is not effective, but $\Symp(2g,\RR)/\{\pm I_g\}$ acts
effectively on $\HH_g$.  An important subgroup of $\Symp(2g,\RR)$ is the
Siegel modular group $\Gamma_g:=\Symp(2g,\ZZ)$ which acts properly
discontinuously on $\HH_g$.

For a function $F$ on $\HH_g$, we introduce the weight-$k$ slash
operation by
\begin{equation}
  \left( F\vert_k
    \gamma\right)(\Omega):=\det(C\Omega+D)^{-k}F(\gamma\cdot\Omega)\,,
\end{equation}
where $k\in \ZZ$ and $\gamma$ is as in \eqref{gamma}.
\begin{df}
  Let $\Gamma$ be a finite index subgroup of $\Gamma_g$ ($g\in \NN$)
  and let $\varrho:\Gamma \rightarrow \CC^\times :=
  \CC\smallsetminus\{0\}$ be a character. A holomorphic function
  $F:\HH_g\rightarrow \CC$ is called a modular form of weight $k$
  ($k\in \ZZ$) with character $\varrho$ if $F$ satisfies
\begin{equation}
  \left( F\vert_k
    \gamma\right)(\Omega)=\varrho(\gamma)F(\Omega),\quad(\forall
  \gamma\in \Gamma)\,,
\end{equation}
(and for $g=1$ holomorphicity conditions at the  cusps of $\Gamma$.)
\end{df}
The space of modular forms is denoted by
${\mathfrak{M}}^g_k(\Gamma,\varrho)$. In particular, we set
${\mathfrak{M}}^g_k(\Gamma) :
={\mathfrak{M}}^g_k(\Gamma,\mathrm{id}_\Gamma)$. The normalized
Eisenstein series $E_k(\tau)$ $(\tau \in \HH_1)$ are defined by
\begin{equation}
  E_k(\tau)=1-\frac{2k}{B_k}\sum_{n=1}^\infty
  \sigma_{k-1}(n)q^n\,,\quad 
q={\bf e}[\tau]\,,\quad (k\in2\NN)\,,
\end{equation}
where $B_k$ are the Bernoulli numbers and $\sigma_k(n)=\sum_{d\mid n}
d^k$.  As is well-known, $\mathop{\oplus}\limits_{k=0}^\infty
{\mathfrak{M}}^1_k(\Gamma_1)\cong\CC[E_4,E_6]$.  Igusa [Igu] determined the
structure of the ring
$\mathop{\oplus}\limits_{k=0}^\infty{\mathfrak{M}}^2_k(\Gamma_2)$ as
\begin{equation}
  \mathop{\oplus}\limits_{k=0}^\infty{\mathfrak{M}}^2_k(\Gamma_2)
  \cong\CC[{\cal E}_4,{\cal E}_6, \chi_{10},\chi_{12},\chi_{35}]^*\,,
\end{equation}
where ${\cal E}_4$ and ${\cal E}_6$ are the genus two Eisenstein
series of weight $4$ and $6$ respectively, while $\chi_{10}$,
$\chi_{12}$ and $\chi_{35}$ are respectively certain cusp forms of
weight $10$, $12$ and $35$. The asterisk is to remind that the cusp
form $\chi_{35}$ is not algebraically independent with $(\chi_{35})^2$
being expressed as a polynomial of the other generators.

Now we turn to the definition of Jacobi forms. The theory of Jacobi
forms with integer weights and integer indices without characters is
well expounded in the monograph of Eichler and Zagier [EZ]. However,
for our purpose, it is convenient to slightly extend the notion of
Jacobi forms so that Jacobi forms of half integral indices with
characters can also be treated. Here we follow the presentation given
in [GN2].

For any pair $[(\lambda,\mu),\kappa]$ and $[(\lambda',\mu'),\kappa']$
with $(\lambda,\mu,\kappa)$, $(\lambda',\mu',\kappa')\in \ZZ^3$,
let  their product be defined by
\begin{equation}
  [(\lambda,\mu),\kappa][(\lambda',\mu'),\kappa']
=[(\lambda+\lambda',\mu+\mu'),
\kappa+\kappa'+\lambda\mu'-\lambda'\mu]\,.
\end{equation}
With this multiplication law the set $
\mathrm{H}_\ZZ:=\{[(\lambda,\mu),\kappa] \mid (\lambda,\mu,\kappa)\in
\ZZ^3\} $ becomes a group and is called the integer Heisenberg group.

For $A=\bigl(
\begin{smallmatrix}
  a&b\\c&d
\end{smallmatrix}
\bigr)\in \Gamma_1$ and $(\lambda,\mu,\kappa)\in \ZZ^3$ consider the
matrix
\begin{equation}
  [A,\lambda,\mu,\kappa]:=\begin{pmatrix}
1 & 0 & 0 & \mu  \\
\lambda  & 1 & \mu  & \kappa  \\
0 & 0 & 1 &  - \lambda  \\
0 & 0 & 0 & 1
\end{pmatrix}
\begin{pmatrix}
a & 0 & b & 0 \\
0 & 1 & 0 & 0 \\
c & 0 & d & 0 \\
0 & 0 & 0 & 1
\end{pmatrix}\in\Gamma_2\,.
\end{equation}
The Jacobi group is the subgroup of $\Gamma_2$ defined by
\begin{equation}
  \Gamma^J:=\{[A,\lambda,\mu,\kappa]\mid \text{$A\in \Gamma_1$ and
    $(\lambda,\mu,\kappa)\in \ZZ^3$}\}\,.
\end{equation}
Since
\begin{equation}
  \begin{split}
 \Gamma^J &\supset \{[A,0,0,0]\mid A\in \Gamma_1\}\cong\Gamma_1,\\[1mm]
\Gamma^J &\vartriangleright \{[I_2,\lambda,\mu,\kappa]
\mid (\lambda,\mu,\kappa)\in \ZZ^3\}\cong\mathrm{H}_\ZZ\,,
\end{split}
\end{equation}
it follows that 
\begin{equation}
  \Gamma^J\cong\Gamma_1\ltimes \mathrm{H}_\ZZ\,.
\end{equation}

Now one can introduce a character of the integer Heisenberg group
$\varrho_\mathrm{H}:\mathrm{H}_\ZZ \rightarrow \{\pm 1\}$ by
\begin{equation}
  \varrho_\mathrm{H}([(\lambda,\mu),\kappa])
=(-1)^{\lambda+\mu+\lambda\mu+\kappa},\quad
  (\lambda,\mu,\kappa)\in \ZZ^3\,.
\end{equation}
Suppose that a character $\varrho_1:\Gamma_1\rightarrow \CC^\times$ is
given.  Then one can construct a character of the Jacobi group
$\varrho_1\times \varrho_\mathrm{H}^l:\Gamma^J\rightarrow \CC^\times$
for any integer $l$ by
\begin{equation}
  (\varrho_1\times
  \varrho_\mathrm{H}^l)([A,\lambda,\mu,\kappa])
=\varrho_1(A)\varrho_\mathrm{H}([(\lambda,\mu),\kappa])^l\,,\quad
  [A,\lambda,\mu,\kappa]\in \Gamma^J\,.
\end{equation}
The Dedekind $\eta$-function is defined by 
\begin{equation}
  \eta(\tau):=q^{\frac{1}{24}}\prod_{n=1}^\infty (1-q^n)\,,\quad
  q={\bf e}[\tau]\,, \quad \tau \in \HH_1\,.
\end{equation}  
$\eta(\tau)^2$ is a weight one modular form of $\Gamma_1$ with a
character. This character is denoted as $\epsilon:\Gamma_1\rightarrow
\CC^\times$. In this paper we always assume that the characters of the
Jacobi group are of the form $\epsilon^a\times
\varrho_\mathrm{H}^b:\Gamma^J\rightarrow \CC^\times$ where $a$ and $b$
are integers.

\begin{df}
  Let $k\in \ZZ$ and $m\in \frac{1}{2}\NN$. A holomorphic function
  $\phi:\HH_1\times\CC\rightarrow \CC$ is called a Jacobi form of
  weight $k$ and index $m$ with a character $\varrho^J:\Gamma^J
  \rightarrow \CC^\times$ if the function $\Phi:\HH_2\rightarrow \CC$
  defined by
  \begin{equation}
   \Phi(\Omega):= p^m\phi(\tau,z)\,,\quad \Omega=\begin{pmatrix}
  \tau&z\\ z&\sigma
      \end{pmatrix}\in \HH_2\,,\quad p={\bf e}[\sigma]\,,
  \end{equation}
  satisfies the  same transformation law under the action of
  $\Gamma^J$ as for the elements of
  $\mathfrak{M}_k^2(\Gamma^J,\varrho^J)$ and if it has an expansion
  of the form
\begin{equation}\label{expansion}
  \phi(\tau,z)=\sum_{\substack{n, r\\ 4mn-r^2\ge
      0}}c(n,r)\,q^n\,y^r\,,\quad q={\bf e}[\tau]\,,\ y={\bf e}[z]\,,
\end{equation}
where $n$ and $r$ run over respectively $\ZZ+\frac{a}{12}$ and
$\ZZ+\frac{b}{2}$ for $\varrho^J=\epsilon^a\times
\varrho_\mathrm{H}^b$ with the restriction $4mn-r^2\ge 0$.
\end{df}
If we replace the condition $4mn-r^2\ge 0$ by $4mn-r^2 > 0$ in the
above, $\phi$ is called a {\it cusp} Jacobi form, while if the
condition $4mn-r^2\ge 0$ is replaced by $n\ge 0$, $\phi$ is called a
{\it weak} Jacobi form.

In the following we frequently use the notation
\begin{equation}
  p={\bf e}[\sigma]\,, \quad q={\bf e}[\tau]\,,\quad  y={\bf e}[z]\,.
\end{equation}

\section{3. Elliptic genera}
Loosely speaking, elliptic genera [Lan1] [Wit] are the equivariant
loop-space indices where equivariance is with respect to the natural
circle action on the loop space.  Physically, they are the extensions
of the Witten index in the sense that they are associated with two
dimensional supersymmetric quantum field theories while the ordinary
Witten index is associated with a supersymmetric quantum mechanics.

In fact, there are several versions of elliptic genera, however what
we shall be concerned with corresponds to the following setting. For a
compact K{\" a}hler manifold $M$, let $E\rightarrow M$ be a fixed
holomorphic vector bundle and let $T\rightarrow M$ be the holomorphic
tangent bundle. Using the splitting principle, the total Chern classes
of $E$ and $T$ can be expressed as
\begin{equation}
 c(E)=\prod_{n=1}^l(1+\alpha_n)\,,\quad c(T)=\prod_{i=1}^d(1+\xi_i)\,,
\end{equation}
where $l:=\rank E$ and $d:=\dim_\CC M$. We assume in the following
$l\ge d$.  The elliptic genus is a function on $\HH_1\times \CC$
defined by
  \begin{equation}\label{egenus}
    Z_E[M](\tau,z):=\int_M
    \prod_{n=1}^lP(\tau,z+\alpha_n)\prod_{i=1}^d\frac{\xi_i}{
      P(\tau,-\xi_i)}\,,
  \end{equation}
where 
\begin{equation}
  P(\tau,z)=\sqrt{-1}q^{\frac{1}{12}}y^{-\frac{1}{2}}
    \prod_{n=1}^\infty(1-yq^{n-1})(1-y^{-1}q^n)\,.
\end{equation}
Notice that we have 
$P(\tau,z)=\vartheta_1(\tau,z)/\eta(\tau)$ by using one of the Jacobi
theta functions.

For any holomorphic vector bundle $F\rightarrow M$ of rank $m$, we
introduce the generating functions for the antisymmetric as well as
symmetric powers of $F$ by
\begin{equation}
  \bigwedge\nolimits_t F=\sum^m_{s=0}t^s(\wedge^s F)\,,\qquad 
 \mathcal{S}_t F=\sum_{s=0}^\infty t^s  (\mathrm{S}^s F)\,.
\end{equation}
Then the  elliptic genus can be expanded as 
\begin{equation}\label{egenusexp}
  \begin{split}
Z_E[M](\tau,z)&=
({ \sqrt{-1}})^{l-d}q^{\frac{l-d}{12}}y^{-\frac{l}{2}}\\[2mm]
&\hspace{1cm}\times \int_M 
\ch \biggl(\mathop{\bigotimes}\limits_{n=1}^\infty
\sideset{}{_{-yq^{n-1}}E} 
\bigwedge\otimes\mathop{\bigotimes}_{n=1}^\infty
\sideset{}{_{-y^{-1}q^{n}}}\bigwedge E^*\\[2mm]
&\hspace{3.5cm}
\otimes\mathop{\bigotimes}_{n=1}^\infty
\mathcal{S}_{q^n}T\otimes\mathop{\bigotimes}_{n=1}^\infty 
\mathcal{S}_{q^n}T^* \biggr )
\td(M)\\[2mm]
&=({ \sqrt{-1}})^{l-d}q^{\frac{l-d}{12}}y^{-\frac{l}{2}}\Big[
\chi_y(E)+O(q)\Big]\,,
\end{split}
\end{equation}
where $E^*$ and $T^*$ are the dual bundles of $E$ and $T$,
respectively and we have set
\begin{equation}
\chi_y(E):=\sum_{s=0}^l(-y)^s\chi(\wedge^s E)\,.
\end{equation}
In the second equality of \eqref{egenusexp}
we have used the Riemann-Roch-Hirzebruch theorem
\begin{equation}
  \chi(E):=\sum_{n=0}^d(-1)^n\dim { H}^n(M,E)=
\int_M\ch(E)\td(M)\,.
\end{equation}
In particular we have
\begin{equation}
  \chi_y(T^*)=\sum_{m,n=0}^d (-1)^{m+n}  h^{m,n} y^m\,,
\end{equation}
where $h^{m,n}:=\dim_\CC H^n(M,\wedge^m T^*)$ are the Hodge numbers of
$M$.  In the sequel we will write $\chi_y(M)$ for $
\chi_y(T^*)$. $\chi_y(M)$ is essentially the $\chi_y$-genus of
Hirzebruch.

The elliptic genus \eqref{egenus} enjoys  nice functional properties if
appropriate topological conditions on $E$ and $T$ are met. The
following summarizes what has been shown in [KYY][KM].
\begin{theorem}
  The elliptic genus $Z_E[M]:\HH_1\times \CC\rightarrow \CC$ is a weak
  Jacobi form of weight $0$ and index $l/2$ with character
  $\varepsilon^{l-d}\times \varrho_\mathrm{H}^l:\Gamma^J\rightarrow
  \CC^\times$ if $c_1(E)=0$ and $\ch_2(E)=\ch_2(T)$.
\end{theorem}
\begin{cor}
  If $M$ is a Calabi-Yau manifold, then the elliptic genus
  $Z_{T^*}[M]:\HH_1\times \CC\rightarrow \CC$ is a weak Jacobi form of
  weight $0$ and index $d/2$ with character
  $\mathrm{id}_{\Gamma_1}\times
  \varrho_\mathrm{H}^d:\Gamma^J\rightarrow \CC^\times$.
\end{cor}
\begin{cor}\label{evenCY}
  If $M$ is an even dimensional Calabi-Yau manifold, then the elliptic
  genus $Z_{T^*}[M]:\HH_1\times \CC\rightarrow \CC$ is a weak Jacobi
  form of weight $0$ and index $d/2$ without character.
\end{cor}

In harmony with this theorem and corollaries, there is a uniform and
concrete procedure [KYY] [KM] to calculate the elliptic genera  of the
Landau-Ginzburg orbifolds corresponding to certain 
realizations of Calabi-Yau manifolds and bundles on them.

In the rest of this paper we shall exclusively be concerned with the
situation of Corollary \ref{evenCY}.  Thus the theory of Jacobi forms
covered in [EZ] is enough.

Instead of $Z_{T^*}[M](\tau,z)$ we will simply write $Z[M](\tau,z)$ or
even $Z(\tau,z)$ when the manifold $M$ is clear from the context.

\section{4. The case of K3 surfaces}
K3 surfaces are the first interesting cases of even dimensional K{\"
  ahler} manifolds of vanishing first Chern classes for which the
elliptic genus is non-trivial. (The elliptic genus of abelian surfaces
vanishes.) So let $M$ be a K3 surface. Then from the knowledge of the
Hodge numbers of K3 surfaces we obtain that
\begin{equation}
  \chi_y(M)=2+20y+2y^2\,.
\end{equation}
Corollary \ref{evenCY} tells us that the elliptic genus $Z(\tau,z)$ is
a weak Jacobi form of weight $0$ and index $1$ with the expansion
\begin{equation}
  Z(\tau,z)=2y^{-1}+20+2y+O(q)\,.
\end{equation}
The first task in this section is to find explicit expressions of
$Z(\tau,z)$.

There are two cusp Jacobi forms of index 1
\begin{equation}
  \begin{split}
    \label{phi1012}
  \phi_{10,1}(\tau,z)&=\frac{E_6(\tau)\,E_{4,1}(\tau,z)
-E_4(\tau)\,E_{6,1}(\tau,z)}{144},\\[1mm]
  \phi_{12,1}(\tau,z)&=\frac{{E_4}(\tau)^2E_{4,1}(\tau,z)
-E_6(\tau)\,E_{6,1}(\tau,z)}{144},
\end{split}
\end{equation}
which are of weight 10 and 12, respectively [EZ] and are the first
coefficients in the Fourier-Jacobi expansions of the Igusa cusp forms
$\chi_{10}$ and $\chi_{12}$, respectively. Here $E_{k,1}(\tau,z)$
$(k\in 2\NN,\ k\ge 4)$ is the Eisenstein-Jacobi series of weight $k$
and index 1 [EZ].  Accordingly there are two weak Jacobi forms of
index one
\begin{equation}\label{tphi1012}
  \tilde\phi_{-2,1}(\tau,z)=\frac{\phi_{10,1}(\tau,z)}{\Delta(\tau)}\,,
\qquad
  \tilde\phi_{0,1}(\tau,z)= \frac{\phi_{12,1}(\tau,z)}{\Delta(\tau)}\,,
\end{equation}
whose weights are respectively $-2$ and $0$. Here we have introduced
the discriminant
\begin{equation}
  \Delta(\tau):=\frac{E_4(\tau)^3-E_6(\tau)^2}{1728}=\eta(\tau)^{24}\,.
\end{equation}
Now comparing the expansions of $Z$ and $\tilde\phi_{0,1}$ and using
the structure theorem \ref{EZFF} mentioned below we find that
\begin{equation}\label{egenusto01}
  Z(\tau,z)=2\tilde\phi_{0,1}(\tau,z)\,.
\end{equation}
This is one way of expressing the elliptic genus of K3 surfaces.

To find another expression, we notice the following identity proved in
[EZ]:
\begin{equation}\label{ratio}
  \frac{\phi_{12,1}(\tau,z)}{\phi_{10,1}(\tau,z)} =
  \frac{\tilde\phi_{0,1}(\tau,z)}{\tilde\phi_{-2,1}(\tau,z)} =
  12\,\wp(\tau,z)\,,
\end{equation}
where the Weierstra{\ss}  $\wp$-function is defined as
\begin{equation}
  \wp(\tau,z) := \frac{1}{(2\pi \sqrt{-1})^2} \left\{ \frac{1}{z^2} +
    \sum_{\substack{\omega\in \ZZ+ \ZZ\tau\\ \omega\ne
        0}}\left(\frac{1}{(z-\omega)^2} -
      \frac{1}{\omega^2}\right)\right\}\,.
\end{equation}
Since we use several properties of the $\wp$-function in this paper we
summarize them below.

The $\wp$-function satisfies the nonlinear differential equation
\begin{equation}\label{perel}
  \begin{split}
  \{D_y\wp(z)\}^2&=4\wp(z)^3-\frac{1}{12}\, E_4\,
  \wp(z)+\frac{1}{216}\, E_6\\[2mm] &=4(\wp(z)-e_1)(\wp(z) -
  e_2)(\wp(z)-e_3)\,,
\end{split}
\end{equation}
where we have introduced the Euler derivative
$D_y:=y\frac{\partial}{\partial y}$.  The roots $e_1$, $e_2$, $e_3$
are related with the half periods
\begin{equation}
  \omega_1=\frac{1}{2},\quad \omega_2=-\frac{\tau+1}{2},\quad
  \omega_3=\frac{\tau}{2}\,,
\end{equation}
by $e_\nu=\wp(\omega_\nu)$ $(\nu=1,2,3)$
and satisfy 
\begin{equation}
  \begin{split}
    e_1+e_2+e_3&=0\,,\\
e_1e_2+e_2e_3+e_3e_1&=-\frac{1}{48}\,E_4\,,\\
e_1e_2e_3&=-\frac{1}{864}\,E_6\,.
  \end{split}
\end{equation}
Furthermore we have useful identities
\begin{align}
  &\sum_{(\lambda,\mu,\nu)\in {\cal S}} (e_\lambda-e_\mu)^2 =
  3({e_1}^2+{e_2}^2+{e_3}^2)=\frac{1}{8}\, E_4\,, \nonumber\\ 
 &\sum_{(\lambda,\mu,\nu)\in {\cal S}}
  (e_\lambda-e_\mu)^2\,e_\nu=-3({e_1}^3+{e_2}^3+{e_3}^3)=\frac{1}{96}\,
  E_6\,, \label{Eines}\\ 
  &\sum_{(\lambda,\mu,\nu)\in {\cal S}}
  (e_\lambda-e_\mu)^2\,{e_\nu}^2={e_1}^4+{e_2}^4+{e_3}^4=
  \frac{1}{1152}\, {E_4}^2\,, \nonumber
\end{align}
where ${\cal S}:=\{(1,2,3),(2,3,1),(3,1,2)\}$.

If we define the theta functions with characteristics by
\begin{equation}
\vartheta\tharg{a}{b}(\tau,z):=\sum_{n\in\ZZ}
\exp\left(\pi \sqrt{-1} \tau (n+a)^2 +2\pi \sqrt{-1} (n+a)(z+b)\right)\,,
\end{equation}
then the Jacobi theta functions are expressed as
\begin{equation}
  \begin{split}
 \vartheta_1(\tau,z)& = 
\vartheta\tharg{\frac{1}{2}}{\frac{1}{2}}(\tau,z)\,,\qquad
\vartheta_2(\tau,z)=\vartheta\tharg{\frac{1}{2}}{0}(\tau,z)\,,\\[1mm]
\vartheta_3(\tau,z)&=\vartheta\tharg{0}{0}(\tau,z)\,,\qquad
\vartheta_4(\tau,z)=\vartheta\tharg{0}{\frac{1}{2}}(\tau,z)\,.   
  \end{split}
\end{equation}
The roots can be written in terms of theta constants as
\begin{align}
  e_1&=-\frac{1}{12}\left\{\vartheta_3(0)^4+\vartheta_4(0)^4\right\}\,,
\nonumber\\
  e_2&=-\frac{1}{12}\left\{
    \vartheta_2(0)^4-\vartheta_4(0)^4\right\}\,,\\
  e_3&=\frac{1}{12}\left\{ \vartheta_2(0)^4+\vartheta_3(0)^4\right\}\,.
\nonumber
\end{align}
We also recall that 
\begin{equation}
  \vartheta_1'(\tau,0)=2\pi\eta(\tau)^3\,,
\end{equation}
where the prime represents the derivative with respect to the second
variable.

We now introduce the function
\begin{equation}\label{Kdef}
  K(\tau,z):=\sqrt{-1}\,\frac{\vartheta_1(\tau,z)}{\eta(\tau)^3}
=2\pi\sqrt{-1}\,\frac{\vartheta_1(\tau,z)}{\vartheta_1'(\tau,0)}\,,
\end{equation}
which plays an important role later in this paper. This function
 is related to the weak Jacobi form $\tilde\phi_{-2,1}$ by
\begin{equation}\label{Kphi}
  \tilde\phi_{-2,1}(\tau,z)=K(\tau,z)^2\,.
\end{equation}

One of the highlights of the theory of Jacobi forms is the following
structure theorem:
\begin{theorem}\label{EZFF}
  {\rm (Eichler-Zagier [EZ], Feingold-Frenkel [FF]).} The bigraded
  ring of all the weak Jacobi forms of even weight is a polynomial
  algebra over $\CC[E_4,E_6]$ generated by $\tilde \phi_{0,1}$ and
  $\tilde \phi_{-2,1}$.
\end{theorem}

\begin{cor} \label{stth} 
  If we assign weights $4$, $6$ and $2$ respectively to $E_4$, $E_6$
  and $\wp$, any weak Jacobi form of weight $2k$ $(k\in \NN_0)$ and
  index $m$ $(m\in \NN_0)$ can be expressed as
\begin{equation*}
  p_{2k+2m}(E_4,E_6,\wp)K^{2m}\,,
\end{equation*}
where $p_{2k+2m}(E_4,E_6,\wp)$ is a weight $2k+2m$ homogeneous
polynomial in $E_4$, $E_6$, $\wp$ and its degree as a polynomial in
$\wp$ is at most $m$.
\end{cor}

Both \eqref{egenusto01} and \eqref{Kphi} are the consequences of
Theorem \ref{EZFF}.  Corollary \ref{stth} is particularly useful for
our later purpose.

Now it trivially follows from \eqref{egenusto01}, \eqref{ratio} and
\eqref{Kphi} that
\begin{equation}\label{massless}
Z(\tau,z)=24\,\wp(\tau,z)\, K(\tau,z)^2\,,
\end{equation}
which is of course in accordance with Corollary \ref{stth}.  The
expression \eqref{massless} has an interesting interpretation as we
shall see.

Since the $\wp$-function is related to the Jacobi theta functions
through the identities [EMOT]
\begin{equation}
  \begin{split}
    \wp(z) &= e_\nu +
    \left(\frac{\vartheta_{\nu+1}(z)}{\vartheta_{\nu+1}(0)}\right)^2\,
    K(z)^{-2},\quad (\nu=1,2,3)\,, \label{petheta}\\ &=
    \frac{1}{3}\sum_{\nu=1}^3\left(
      \frac{\vartheta_{\nu+1}(z)}{\vartheta_{\nu+1}(0)}\right)^2\,
    K(z)^{-2}\,,
\end{split}
\end{equation}
the elliptic genus can also be expressed as\footnote{\, The formula
  given in [KYY] corresponds to the choice $\nu=2$.}
\begin{equation}
  Z(\tau,z) = 24\left\{ \left(
      \frac{\vartheta_{\nu+1}(z)}{\vartheta_{\nu+1}(0)}\right)^2+
    e_\nu\, K(z)^2\right\},\quad (\nu=1,2,3)\,,
\end{equation}
or
\begin{equation}
  Z(\tau,z) = 8\sum_{\nu=1}^3\left(
    \frac{\vartheta_{\nu+1}(z)}{\vartheta_{\nu+1}(0)}\right)^2\,.
\end{equation}
These formulas are related to the Kummer surface which is realized by
resolving the singularities of the orbifold $\CC^2/\ZZ_2$. (For
instance, {\it cf.\/} [Wal] [EOTY].)

As it easily follows from \eqref{phi1012}, \eqref{ratio} and
\eqref{Kphi} we have the formulas
\begin{equation}
  \begin{split}
E_{4,1}&=\left(E_4\, \wp-\frac{1}{12}\, E_6\right)K^2\,,\\
E_{6,1}&=\left(E_6\, \wp-\frac{1}{12}\, {E_4}^2\right)K^2\,,
\end{split}
\end{equation}
which are again in accordance with Corollary \ref{stth}.  There is
another way of expressing the Eisenstein-Jacobi series $E_{4,1}$ and
$E_{6,1}$ which reads
\begin{equation}\label{Epeth}
  \begin{split}
  E_{4,1}&=\frac{1}{2}\sum_{\nu=1}^3 \vartheta_{\nu+1}(z)^2
  \vartheta_{\nu+1}(0)^6\,,\\[2mm]
  E_{6,1}&=6\sum_{\nu=1}^3\wp(z+\omega_\nu)\vartheta_{\nu+1}(z)^2
  \vartheta_{\nu+1}(0)^6\,.
\end{split}
\end{equation}
It is not difficult to check that the two expressions indeed coincide by
use of formulas
\begin{equation}
\begin{split}
  \vartheta_{\nu+1}(0)^8 &= 16 (e_\lambda-e_\mu)^2,\quad
  (\lambda,\mu,\nu)\in {\cal S}\,,\\ 
    \wp(z+\omega_\nu) &=
  e_\nu+\frac{(e_\nu-e_\lambda)(e_\nu-e_\mu)}{\wp(z)-e_\nu}, \quad
  (\lambda,\mu,\nu)\in {\cal S}\,,
\end{split}
\end{equation}
as well as \eqref{petheta} and \eqref{Eines}.  The second expressions
\eqref{Epeth} appear in certain compactifications of heterotic string.

The theta functions associated with the integrable representations at
level $m$ $(m\in \NN)$ of the affine Lie algebra $A_1^{(1)}$ are
defined by
\begin{equation}
  \theta_{\mu,m}(\tau,z):= \sum_{\substack{r\in\ZZ\\ r\equiv \mu \bmod
      {2m}}} q^{\frac{r^2}{4m}}y^{\frac{r}{2}}\,, \quad \mu\in \ZZ
  \mod 2m\ZZ\,.
\end{equation}
Any (weak) Jacobi form $\phi(\tau,z)$ of index $m$ can be expressed  in
terms of these theta functions as [EZ, \S 5]
\begin{equation}\label{expA1}
  \phi(\tau,z)=\sum_{\mu(\bmod 2m)}h_\mu(\tau) \theta_{\mu,m}(\tau,2z)\,,
\end{equation}
for some set of functions $h_\mu(\tau)$.

 The elliptic genus of K3 surfaces can be expanded as
\begin{equation}\label{egenuscoeff}
  Z(\tau,z)=\sum_{n\in \NN_0,\ r\in \ZZ} c(4n-r^2)q^ny^r\,,
\end{equation}
where we used the fact that the coefficients depend only on the
combination $4n-r^2$.  By defining
\begin{equation}
  h_0(\tau):=\sum_{N \equiv 0 \mod 4}c(N)q^{\frac{N}{4}}\,,\qquad
 h_1(\tau):=\sum_{N \equiv -1 \mod 4}c(N)q^{\frac{N}{4}}\,,
\end{equation}
the elliptic genus  can thus be written as
\begin{equation}
  Z(\tau,z)=h_0(\tau)\theta_{0,1}(\tau,2z)
  +h_1(\tau)\theta_{1,1}(\tau,2z)\,.
\end{equation}
The $q$-expansions of the functions $h_0(\tau)$ and $h_1(\tau)$ are given
by
\begin{equation}
  \begin{split}
    h_0(\tau)&=20 + 216\,{q} + 1616\,{q}^{2} + 8032\,{q}^{3} +
    33048\,{ q}^{4} + \cdots\,,\\ 
    q^{\frac{1}{4}}h_1(\tau)&=2 - 128\,{q} - 1026\,{q}^{2} -
    5504\,{q}^{3} - 23550\,{q}^{4}-\cdots\,.
\end{split}
\end{equation}
Note in particular that 
\begin{equation}
  c(0)=20\,,\qquad c(-1)=2\,,\qquad \text{$c(N)=0$ if $N <-1$}\,.
\end{equation}
There are explicit expressions of $h_0$
and $h_1$  in terms of theta constants:
\begin{equation}
  \begin{split}
    h_0 &= \frac{10\,\theta_{0,1}(0)^4\,\theta_{1,1}(0) -
      2\,\theta_{1,1}(0)^5}{\eta^6}\,,\\ 
    h_1&=\frac{2\,\theta_{0,1}(0)^5 -
      10\,\theta_{0,1}(0)\,\theta_{1,1}(0)^4}{\eta^6}\,.
\end{split}
\end{equation}
This can be demonstrated by applying the formulas
\begin{equation}
  \begin{split}
    \vartheta_1(z)^2&= 
 \theta_{1,1}(0)\theta_{0,1}(2z)-\theta_{0,1}(0) \theta_{1,1}(2z)\,,  \\
\vartheta_2(z)^2&= 
 \theta_{1,1}(0)\theta_{0,1}(2z)+\theta_{0,1}(0) \theta_{1,1}(2z)\,,  \\
\vartheta_3(z)^2&= 
  \theta_{0,1}(0)\theta_{0,1}(2z)+\theta_{1,1}(0) \theta_{1,1}(2z)\,, \\
\vartheta_4(z)^2&=   
\theta_{0,1}(0)\theta_{0,1}(2z)-\theta_{1,1}(0) \theta_{1,1}(2z)\,.
\end{split}
\end{equation}

\section{5. Product formulas of Borcherds, Gritsenko and Nikulin}
Let $V$ be a $2+n$-dimensional real vector space and let $(\ ,\ ):
V\times V\rightarrow \RR$ be a non-degenerate bilinear form of
signature $(2,n)$ on $V$. The orthogonal group $O(V)$ consists of four
connected components. The index two subgroup consisting of the
elements of spinor norm 1 in $O(V)$ is denoted as $O^+(V)$. The
connected component that contains the identity is denoted by $SO^+(V)$
so that $SO^+(V)\cong O^+(V)/\{\pm I_{2+n}\}$. Consider an orthogonal
decomposition
\begin{equation}\label{decomposition}
  V=V_+\oplus V_-\,,
\end{equation}
such that $(\ ,\ )\vert_{V_+}$ is positive definite and $(\ ,\ )
\vert_{V_-}$ is negative definite. Let $\mathcal{K}$ be the subgroup
of $SO^+(V)$ which respects this decomposition. $\mathcal{K}$ is a
maximal compact subgroup of $SO^+(V)$. Thus inequivalent
decompositions \eqref{decomposition} are parametrized by
$SO^+(V)/\mathcal{K}$ which is a hermitian symmetric space of type IV
in the classification of E. Cartan. Let $\mathcal{D}$ be one of the
two connected components of
\begin{equation}\label{domain}
  \{ [w]\in \PP(V\otimes \CC) \mid
  \text{$(w,w)=0$ and $(w,\bar w)>0$}\}\,,
 \end{equation}
 where $[w]$ is a line corresponding to $w\in V\otimes
 \CC\smallsetminus \{0\} $.  There exists an isomorphism
 $SO^+(V)/\mathcal{K}\cong \mathcal{D}$. To see this, let $w_1=\Rea w$
 and $w_2=\Ima w$ for $[w]\in \mathcal{D}$.  Then the conditions in
 \eqref{domain} are equivalent to
\begin{equation}
  \text{$(w_1,w_1)=(w_2,w_2)>0$ and
    $(w_1,w_2)=0$}\,.
\end{equation}
Thus $[w]\in \mathcal{D}$ determines a real two dimensional (positive)
vector space $V_+^{[w]}$ spanned by the (oriented) normalized
orthogonal basis $\{\hat w_1,\hat w_2\}$ where
\begin{equation}
  \hat w_i:=\frac{1}{\sqrt{(w_i,w_i)}}\,w_i =
  \sqrt{\frac{2}{(w,\bar w)}}\,w_i\,,\quad (i=1,2)\,.
\end{equation}

To be specific, 
assume in the following that $V$ is spanned by
five vectors $e_1$, $e_2$, $f_1$, $f_2$ and $\delta$
with their only non-vanishing inner products given by
\begin{equation}
  (e_1,e_2)=(e_2,e_1)=1=(f_1,f_2)=(f_2,f_1)\,,\quad
     (\delta,\delta)=-\frac{1}{2}\,.
\end{equation}
Then consider the lattice 
\begin{equation}
  L:=H^{(1)}\oplus H^{(2)}\oplus \ZZ\delta\,,
\end{equation}
where $H^{(1)}=\ZZ e_1+\ZZ e_2$ and $H^{(2)}=\ZZ f_1+\ZZ f_2$,
together with its sublattice
\begin{equation}
  \Lambda:=H^{(2)}\oplus \ZZ\delta\,.
\end{equation}
Obviously $V=L\otimes \RR$ has signature $(2,3)$.
 The cone $\{x\in \Lambda\otimes\RR \mid
(x,x)>0\}$ consists of two connected components. We fix one of them
(the future light-cone) and denote it by $C^+(\Lambda)$.
Then the {\it tube domain\/}
\begin{equation}
  \mathcal{H}:=\Lambda\otimes\RR+\sqrt{-1}C^+(\Lambda)\subset
  \Lambda\otimes\CC\,,
\end{equation} gives a realization of $\mathcal{D}$ with 
the isomorphism $\mathcal{H}\xrightarrow{\cong}\mathcal{D}$ given by
\begin{equation}\label{iso1}
   \zeta\in
  \mathcal{H}   \longrightarrow [w]\in\mathcal{D}\,,\quad
  w=e_1-\frac{(\zeta,\zeta)}{2}e_2+\zeta\,.
\end{equation}
In the present case, we have a further  isomorphism
$\HH_2\xrightarrow{\cong} \mathcal{H}$ given by
\begin{equation}\label{iso2}
\Omega=
  \begin{pmatrix}
   \tau&z\\ z&\sigma 
  \end{pmatrix}
  \in \HH_2 \longrightarrow \zeta=\tau f_1+\sigma f_2+2z\delta\in
  \mathcal{H}\,.
\end{equation}
In what follows we will freely use these isomorphisms. Thus, for
instance we have
\begin{equation}
  (w,\bar w)=2(\Ima(\zeta),\Ima(\zeta))=4\det\Ima(\Omega)>0\,.
\end{equation}

The symplectic group $\Symp(4,\RR)$ acts on $\HH_2$ while the group
$O^+(V)$ acts on the domain $\mathcal{D}$.  We have an isomorphism
$\Symp(4,\RR)/\{\pm I_4\}\cong SO^+(V)$ compatible with the isomorphism
$\HH_2\cong\mathcal{D}$. If we define $O^+(L)=O(L)\cap O^+(V)$ and
$SO^+(L)=O(L)\cap SO^+(V)$ where $O(L)$ is the automorphism group of
$L$, then there exists an isomorphism $\Gamma_2/\{\pm I_4\}\cong
SO^+(L)=O^+(L)/\{\pm I_5\}$. (An explicit demonstration of this for
the dual lattice $L^*$ can be found in [GN1].)  Hence one can regard
Siegel modular forms on $\HH_2$ with respect to (a finite index
subgroup of) $\Gamma_2$ as automorphic forms on the type IV domain
$\mathcal{D}$ with respect to (a finite index subgroup of) $O^+(L)$.

For the product of all the genus two theta functions with even
characteristics denoted as $\Delta_5(\Omega)$, Gritsenko and Nikulin
[GN1] proved an infinite product formula
\begin{equation}\label{delta5}
  \Delta_5(\Omega) =p^{\frac{1}{2}}q^{\frac{1}{2}}y^{\frac{1}{2}}
  \prod\limits_{(k,l,r)>0} (1- p^kq^ly^r)^{c(4kl-r^2)/2}\,,
\end{equation}
where $p={\bf e}[\sigma]$, $q={\bf e}[\tau]$ and $y={\bf e}[z]$ as
before and $c(N)$ are the coefficients appearing in \eqref{egenuscoeff}.
The product in \eqref{delta5} is taken over all triplets of integers
$(k,l,r)$ satisfying the conditions $(k,l,r)>0$ where
\begin{equation*}
  (k,l,r)>0\iff
  \begin{aligned}[t]
    \text{either}&\quad k>0,\ l\ge 0\,,\\
    \text{or}&\quad  k\ge0,\ l> 0\,,\\
    \text{or}&\quad k=l=0,\ r<0\,.
  \end{aligned}
\end{equation*}
Since the relation $\chi_{10}(\Omega)=\Delta_5(\Omega)^2$ holds with a
suitable choice of the multiplicative constant, we have an infinite
product representation of the Igusa cusp form of weight ten:
\begin{equation}
  \chi_{10}(\Omega)=
pqy\prod\limits_{(k,l,r)>0} (1- p^kq^ly^r)^{c(4kl-r^2)}\,.
\end{equation}

In order to prove this infinite product representation Gritsenko and
Nikulin used the lifting procedure introduced by Borcherds [Bor1].
The Borcherds-lifting makes use of the Hecke operators for the Jacobi
group as essential ingredients.

The Hecke operators $V_\ell$ $(l\in \NN)$ on a function
$\phi:\HH_1\times \CC\rightarrow \CC$ is defined [EZ] by
\begin{multline}
  \left( \phi\vert_{k,m}V_\ell\right) (\tau,z):=\\[2mm] \ell^{k-1}
\sum_{\bigl(
  \begin{smallmatrix}
    a&b\\c&d
  \end{smallmatrix}\bigr) \in \Gamma_1\backslash\M_\ell} 
(c\tau+d)^{-k} {\bf e}\left[ \frac{-\ell m cz^2}{c\tau+d}\right] 
\phi \left( \frac{a\tau+b}{c\tau+d},\frac{\ell z}{c\tau+d}\right),
\end{multline}
where $\M_\ell:=\{ A\in \M(2,\ZZ) \mid \det A=\ell\}$.  By choosing
the standard set of representatives we may take [Apo]
\begin{equation}\label{hermite}
\Gamma_1\backslash \M_\ell=\left\{
  \begin{pmatrix}
    a&b\\ 0&d
  \end{pmatrix}\Big\vert\ 
  \text{$a,d \in \NN$, $ ad=\ell$ and $b=0,1,\ldots, d-1$. }\right\}\,.
\end{equation}
Therefore we have more explicitly
\begin{equation}
  \left( \phi\vert_{k,m}V_\ell\right) (\tau,z) =
  \ell^{k-1}\sum_{ad=\ell} \sum_{b=0}^{d-1} d^{-k}
  \phi\left(\frac{a\tau+b}{d},az\right).
\end{equation}

Using these Hecke operators, a straightforward calculation leads to
the identity
\begin{equation}\label{prtoH}
  \prod_{\substack{k>0 \\ \ (k,l,r)>0}}(1-p^kq^ly^r)^{c(4kl-r^2)} =
  \exp\left(-\sum_{\ell=1}^\infty p^\ell (Z\vert_{0,1}
    V_\ell)(\tau,z)\right)\,.
\end{equation}
Consequently we obtain
\begin{equation}\label{borcherds}
  \chi_{10}(\Omega) =p \psi(\tau,z) \exp\left(-\sum_{\ell=1}^\infty
    p^\ell (Z\vert_{0,1} V_\ell)(\tau,z)\right)\,,
\end{equation}
where we have set
\begin{equation}
  \psi(\tau,z):=qy \prod\limits_{(0,l,r)>0} (1 - q^ly^r)^{c(-r^2)}\,,
\end{equation}
which is a Jacobi form of weight $c(0)/2=10$ and index $1$ as is clear
from an alternative expression
\begin{equation}\label{psiexp}
  \psi(\tau,z)=\Delta(\tau)K(\tau,z)^2\,.
\end{equation}
The RHS of \eqref{borcherds} should be compared with the equation
given on p.192 in [Bor1].  The factor $\psi(\tau,z)$ (or its square
root for $\Delta_5$) corresponds to the ``denominator'' of the affine
vector system in [Bor1].

We will see in the following that the decomposition into factors
\eqref{borcherds} possesses some interesting interpretations.

\section{6. Modular integral and Igusa cusp form}
For a given $[w]\in \mathcal{D}$, let $V_-^{[w]}$ be the orthogonal
complement of $V_+^{[w]}$ in $V$ so that $V=V_+^{[w]}\oplus
V_-^{[w]}$. For each $\lambda\in L$, let $\lambda_\pm^{[w]}$ denote
the projection of $\lambda$ onto $V_\pm^{[w]}$.  {}From the obvious
relation $\lambda_+^{[w]}=\sum_{i=1}^2(\lambda,\hat w_i)\hat w_i$ it
follows that
\begin{equation}
  \left(\lambda_+^{[w]},\lambda_+^{[w]}\right) 
=\frac{2\abs{(\lambda,w)}^2}{(w,\bar w)}\,.
\end{equation}
We  divide the lattice  $L$ into two sublattices 
\begin{equation}
  L_0 := H^{(1)}\oplus H^{(2)} \oplus 2\ZZ\delta \quad \text{and}
  \quad L_1:=L_0+\delta\,,
\end{equation}
so that $L=L_0\cup L_1$.  

For a fixed $\Omega=\bigl(
\begin{smallmatrix}
  \tau&z\\z&\sigma
\end{smallmatrix}\bigr)\in \HH_2$  
we will consider the  integral
\begin{equation}\label{modintegral}
  \mathcal{I}(\Omega)= \frac{1}{2} 
\int\limits_{\mathrm{P} \Gamma_1 \backslash\HH_1} 
\frac{d^2\rho}{ \Ima(\rho)} \left(
    h_0(\rho)\sum_{\lambda\in L_0}+h_1(\rho)\sum_{\lambda\in
      L_1}\right)R(\rho,\bar\rho,\lambda,[w])\,,
\end{equation}
where $d^2\rho=d\Rea(\rho)\, d\Ima(\rho)$,
$\mathrm{P}\Gamma_1:=\Gamma_1/\{\pm I_2\}$ and
\begin{equation}
  \begin{split}
    R(\rho,\bar\rho,\lambda,[w]):=& \mathbf{e}\left[ - \rho
      \frac{1}{2}\left(\lambda_-^{[w]},\lambda_-^{[w]}\right) -
      \bar\rho \frac{1}{2}\left(\lambda_+^{[w]},\lambda_+^{[w]}\right)
\right]\\[1mm]  
=&\mathbf{e}\left[-\frac{1}{2}(\lambda,\lambda)\rho\right]
\exp\left(-2\pi\Ima(\rho) \left( \lambda_+^{[w]},\lambda_+^{[w]}
 \right)\right)\,,
  \end{split}
\end{equation}
It should be understood that we are here implicitly using the
isomorphisms given by \eqref{iso1} and \eqref{iso2}. As a matter of
fact, the integral \eqref{modintegral} is divergent and we need a
regularization for it to be meaningful.

Similar integrals have appeared in several physical contexts of string
theory. Dixon, Kaplunovsky and Louis [DKL] were the first to
explicitly evaluate this type of integral. Their method was extended
in the work of Harvey and Moore [HM] where the connections to infinite
products {\it {\` a} la\/} Borcherds [Bor1] were pointed
out.   A further extension can  be found in [Bor2]. 

The particular case of
\eqref{modintegral} was treated in [Kaw1]. The upshot of the
calculation is
\begin{theorem}
  The integral \eqref{modintegral} with an appropriate regularization 
is equal to 
\begin{equation}
  -\log\big( Y^{10}\abs{\chi_{10}(\Omega)}^2\big)\,,
\end{equation}
where $Y:=\det\Ima(\Omega)$.
\end{theorem}

Below we sketch the outline of the proof.  First parametrize
$\lambda\in L$ in terms of five integers $m_1$, $m_2$, $n_1$, $n_2$
and $r$ as
\begin{equation}
  \lambda=-m_1e_1+m_2e_2+n_1f_1+n_2f_2-r\delta\,.
\end{equation}
Then, we find that
\begin{multline}
  R(\rho,\bar\rho,\lambda,[w])= {\bf e}[\rho (m_1m_2-n_1n_2) ]{\bf
    e}[\rho r^2/4]\\[1mm] \times \exp\left(-\frac{\pi\Ima(\rho)}{Y}\,
    \big\lvert
    m_2+n_2\tau+n_1\sigma+m_1(\tau\sigma-z^2)+rz\big\rvert^2\right)\,.
\end{multline}
Substituting this into \eqref{modintegral} and performing the Poisson
resummations over $m_2$ and $n_2$, we arrive at the result
\begin{equation}\label{poisson}
  \begin{split}
    \mathcal{I}(\Omega)&=
    \frac{Y}{2\Ima(\tau)}\int\limits_{\mathrm{P}\Gamma_1\backslash\HH_1}
    \frac{d^2\rho}{ \Ima(\rho)^2}\sum_{G \in \M(2,\ZZ)}
    \left(h_0(\rho)\sum_{r\in 2 \ZZ}+h_1(\rho)\sum_{r \in 2 \ZZ+1}
    \right)\\ &\hspace{4.5cm}\times{\bf e}[\rho r^2/4]{\bf e}[-\sigma
    \det G]\exp(F[G])\\ &=
    \frac{Y}{2\Ima(\tau)}\int\limits_{\mathrm{P}\Gamma_1\backslash\HH_1}
    \frac{d^2\rho}{\Ima(\rho)^2} \\ &\qquad\times \sum_{G \in
      \M(2,\ZZ)} \sum_{N,r\in\ZZ} c(N){\bf e}[\rho(N+ r^2)/4]{\bf
      e}[-\sigma\det G]\exp(F[G])\,,
    \end{split}
\end{equation}
where $F[G]$ assumes a complicated expression which I refrain from
writing down. Notice that we have extended the definition of $c(N)$ by
assigning $c(N)=0$ if $N \equiv 1\pmod 4$ or $N \equiv 2 \pmod 4$. The
crucial point is that the right multiplication of any element of
$\mathrm{P}\Gamma_1$ on $G\in \M(2,\ZZ)$ can be absorbed into the
action of the same element of $\mathrm{P}\Gamma_1$ on $\rho$. Actually
this is not manifest in \eqref{poisson} but can be seen by using the
modular properties of $h_0$ and $h_1$. Thus rather than summing over
$\M(2,\ZZ)$ and integrating over the fundamental region
$\mathrm{P}\Gamma_1\backslash\HH_1$, one may, by restricting oneself
to the sum over the representatives in the orbit decomposition of
$\M(2,\ZZ)$ with respect the right action of $\mathrm{P}\Gamma_1$,
integrate over more affordable regions larger than the fundamental
region.

Explicitly the orbit decomposition is given by
\begin{equation}\label{012}
  \M(2,\ZZ)=\mathcal{O}_0\, \mathrm{P}\Gamma_1
\cup\mathcal{O}_1\, \mathrm{P}\Gamma_1 
\cup\mathcal{O}_2\, \mathrm{P}\Gamma_1\,,
\end{equation}
where
\begin{equation}
  \begin{split}
  &\mathcal{O}_0=\left\{
    \begin{pmatrix}
    0&0\\0&0
    \end{pmatrix}
\right\}\,,\\[2mm]
&\mathcal{O}_1=\left\{ \begin{pmatrix}
    0&m\\0&n
    \end{pmatrix} \Big\vert\ m,n \in \ZZ,\  
(m,n)\not= (0,0)\right\}\,,\\[2mm]
  &\mathcal{O}_2=\bigcup_{\ell \in
    \NN}\pm\M_\ell/\Gamma_1 \cup \bigcup_{\ell \in
    \NN}\pm Q(\M_\ell/\Gamma_1)\,,\quad Q=
\begin{pmatrix}
  1&0\\0&-1
\end{pmatrix}\,,
  \end{split}
\end{equation}
with
\begin{equation}
  \M_\ell/\Gamma_1 =\left\{
  \begin{pmatrix}
    a&b\\ 0&d
  \end{pmatrix}\Big\vert\ 
  \text{$a,d \in \NN$, $ ad=\ell$ and $b=0,1,\ldots, a-1$. }\right\}\,.
\end{equation}
 Accordingly one may consider the decomposition
\begin{equation}
  \mathcal{I}(\Omega)=\mathcal{I}_0(\Omega)+\mathcal{I}_1(\Omega)
+\mathcal{I}_2(\Omega)\,.
\end{equation}

Before evaluating each contribution in this, some intuitive remarks
are in order.  We started from \eqref{modintegral} and reached
\eqref{poisson} by the Poisson resummations. Physically this means
switching from the Hamiltonian picture to the Lagrangian or the path
integral picture of a sigma model with a single Wilson line whose
worldsheet is the elliptic curve $E_\rho:=\CC/(\ZZ+ \rho \ZZ)$
and the target space is again the elliptic curve $E_\tau:=\CC/(\ZZ+
\tau \ZZ)$. The target elliptic curve $E_\tau$ is further parametrized
by the complexified K{\" a}heler parameter $\sigma$. Thus the factor
${\bf e}[-\sigma \det G]$ appearing in \eqref{poisson} can be regarded
as the usual one counting the instanton number.

The contribution $\mathcal{I}_2(\Omega)$ corresponds to the non-zero
degree maps whose images wrap properly the target elliptic curve
possibly many times.  One may intuitively expect that if the area of
the image is very large, {\it i.e.} the image wraps the target
elliptic curve large number of times, its contribution is
exponentially suppressed and $\mathcal{I}_2(\Omega)$ is not divergent.
$\mathcal{I}_0(\Omega)$ is associated with the map whose image shrinks
to a point in $E_\tau$.  This either should not give rise to a
divergence. On the other hand, $\mathcal{I}_1(\Omega)$ corresponds to
the maps whose images degenerate to non-trivial 1-cycles in
$E_\tau$. These cycles can be arbitrarily long and wrap the basis of
$H_1(E_\tau,\ZZ)$ infinitely many number of times without any costs
since they have zero areas. Consequently they can be potential sources
of divergence.

Now we turn to the actual calculations. Since $\mathcal{O}_0\,
\mathrm{P}\Gamma_1=\mathcal{O}_0$, we have to integrate over the
fundamental region to calculate $\mathcal{I}_0(\Omega)$. However this
is not difficult and we obtain that
\begin{equation}
  \mathcal{I}_0(\Omega)=(c(0)+2c(-1))\frac{\pi Y}{6\Ima(\tau)}\,.
\end{equation}
As for $\mathcal{I}_2(\Omega)$, we note that $\mathrm{P}\Gamma_1$ acts
effectively on each element of $\mathcal{O}_2$. Thus we can extend the
integral region from the fundamental region to the whole upper plane
$\HH_1$.  The calculation is more or
less straightforward and the result reads
\begin{equation}
  \mathcal{I}_2(\Omega)=-\log\Bigg( Y^{10}\, \Bigg\lvert
  \prod_{\substack{k>0 \\ \ 
      (k,l,r)>0}}(1-p^kq^ly^r)^{c(4kl-r^2)}\Bigg\rvert^2\,\Bigg)\,.
\end{equation}
The expression inside the absolute value is precisely the one related
to the Hecke operators in the previous section. This should not be too
surprising since determining a holomorphic map of degree $\ell$ from
$E_\rho$ to $E_\tau$ is equivalent to determining a sublattice of
index $\ell$ in the target lattice $\ZZ + \ZZ\tau\subset \CC$ as the
image of the worldsheet lattice $\ZZ+ \ZZ\rho$.  However the set of
all the sublattices of index $\ell$ in $\ZZ+ \ZZ\tau$ is isomorphic to
$\Gamma_1\backslash \M_\ell$ and the Hecke operator $V_\ell$
associates with the lattice $\ZZ+ \ZZ\tau$ the sublattices of index
$\ell$ in $\ZZ+ \ZZ\tau$, with multiplicity one.

Now we turn to the subtle case of $\mathcal{I}_1(\Omega)$.
Notice first that each element of $\mathcal{O}_1$ is fixed by the right
action of $\{\bigl(
\begin{smallmatrix}
  1 & k\\
  0& 1\\
\end{smallmatrix}\bigr) \mid k \in \ZZ\}\subset \mathrm{P}\Gamma_1$.
Thus the integral region can be extended only to the strip of width
one in $\HH_1$.  Furthermore, as mentioned in the above,
$\mathcal{I}_1(\Omega)$ must be regularized somehow.  As a regularized
expression of $\mathcal{I}_1(\Omega)$, we adopt, for a complex number
$s$ with $\Rea(s)>1$, the integral
\begin{equation}\label{regdeg}
  \begin{split}
      \mathcal{I}_1(\Omega,s)&=\frac{Y}{2\Ima(\tau)}\int_{-1/2}^{1/2}
d\Rea(\rho)\int^\infty_0\frac{d\Ima(\rho)}{\Ima(\rho)^{1+s}}\\
&\hspace{2cm}\times\sum_{N,r\in\ZZ} 
c(N){\bf e}[\rho(N+ r^2)/4]\sideset{}{'}\sum_{m,n}\exp\left(F\left[\bigl(
    \begin{smallmatrix}
      0&m\\0&n
    \end{smallmatrix}\bigr)
     \right]\right)\\[2mm]
&=\frac{Y}{2\Ima(\tau)}\int^\infty_0\frac{d\Ima(\rho)}{\Ima(\rho)^{1+s}}
\sum_{r=0,\pm 1} c(-r^2)\sideset{}{'}\sum_{m,n}\exp\left(F\left[\bigl(
    \begin{smallmatrix}
      0&m\\0&n
    \end{smallmatrix}\bigr)
     \right]\right)\,,
 \end{split}
\end{equation}
where $\sideset{}{'}\sum_{m,n}$ means taking the sum over all pair of
integers $(m,n)$ except for $(m,n)=(0,0)$ and
\begin{equation}
  F\left[\bigl(
    \begin{smallmatrix}
      0&m\\0&n
    \end{smallmatrix}\bigr)
     \right]=-\frac{\pi Y}{\Ima(\rho)\Ima(\tau)^2}\abs{m+n\tau}^2
+\frac{2\pi\sqrt{-1}r}{\Ima(\tau)}\Ima(z(m+n\bar\tau))\,.
\end{equation}

To rewrite this expression we need to recall the definitions of some
well-known functions in number theory. We introduce two kinds of them
known as the Epstein zeta functions or alternatively as the
non-holomorphic Eisenstein series.  For $\tau\in\HH_1$ and $s\in \CC$
with $\Rea(s)>1$, the first zeta function is defined by
\begin{equation}
  E(\tau,s)=\sideset{}{'}\sum_{m,n} 
\frac{\Ima(\tau)^s}{\abs{m+n\tau}^{2s}}\,.
\end{equation}
Similarly, for $\tau\in\HH_1$ and $s\in \CC$ with $\Rea(s)>1$, the
second zeta function is defined for real numbers $\alpha$ and $\beta$
which are not both integers by
\begin{equation}
  E(\alpha,\beta\,;\,\tau,s)=\sideset{}{'}\sum_{m,n}{\bf
    e}[m\alpha+n\beta]\frac{\Ima(\tau)^s}{\abs{m+n\tau}^{2s}}\,.
\end{equation}
Now exchanging the sums and the integral in the last expression of
\eqref{regdeg}, which is possible by virtue of our regularization, and
recalling the definition of the Gamma function
$\Gamma(s)=\int^\infty_0dt\, t^{s-1}e^{-t}$, $(\Rea(s)>1)$ we obtain
that
\begin{equation}
  \mathcal{I}_1(\Omega,s) = \frac{\Gamma(s)}{2\pi}\left(
    \frac{\Ima(\tau)}{\pi Y}\right)^{s-1} \big[ c(0)E(\tau,s) + 2c(-1)
  E(\alpha,\beta\,;\,\tau,s)\big]\,,
\end{equation}
with 
\begin{equation}
  \alpha=\frac{\Ima(z)}{\Ima(\tau)}\,,\qquad
\beta=\frac{\Ima(z\bar\tau)}{\Ima(\tau)}\,.
\end{equation}
In order to evaluate this expression near $s=1$ we recall the
Kronecker limit formulas [Sie][Lan2].  \newtheorem*{K1}{First limit
  formula} \newtheorem*{K2}{Second limit formula}
\begin{K1}
  $E(\tau,s)$ can be analytically continued into a function regular
  for $\Rea (s)>\frac{1}{2}$ except for a simple pole at $s=1$. It has
  the expansion
\begin{equation}
  E(\tau,s)=\frac{\pi}{s-1} + 2\pi\left(\gamma_E - \log 2 -
    \log\sqrt{\Ima(\tau)}\abs{\eta(\tau)}^2 \right) + O(s-1)\,,
\end{equation}
where $\gamma_E= \lim\limits_{n\rightarrow\infty}\left(
  1+\frac{1}{2}+\cdots+\frac{1}{n} - \log n \right)$ is Euler's
constant.
\end{K1}
\begin{K2}
  $E(\alpha,\beta\,;\,\tau,s)$ can be  analytically continued into a
  function regular for $\Rea(s)>\frac{1}{2}$ and
\begin{equation}
  E(\alpha,\beta \,;\, \tau,1)
=-\pi\log\left\lvert \frac{\vartheta_1(\tau,\alpha\tau-\beta)}{\eta(\tau)}
{\bf e}[\tau\alpha^2/2]\right\rvert^2\,.
\end{equation}
\end{K2}
Thus analytically continuing $\mathcal{I}_1(\Omega,s)$ and using
$\Gamma(s)=1-\gamma_E(s-1)+O((s-1)^2)$, we obtain the expansion
\begin{multline}
  \mathcal{I}_1(\Omega,s) = c(0)\left(\frac{1}{2(s-1)} +
    \frac{\gamma_E}{2}- \frac{1}{2} \log(4\pi) - \log\sqrt{Y} -
    \log\abs{\eta(\tau)}^2\right)\\ + c(-1) \left( - \log \left\lvert
      \frac{\vartheta_1(\tau,z)}{\eta(\tau)}\right\rvert^2 +
    2\pi\frac{\Ima(z)^2}{\Ima(\tau)} \right) + O(s-1)\,.
\end{multline}
Using this the ``degree zero'' part amounts to
\begin{equation}
  \begin{split}
     \mathcal{I}_0(\Omega)+\mathcal{I}_1^\mathrm{reg}(\Omega)     &=-\log
    \big(Y^{10}\left\lvert{\bf
        e}[\sigma]\eta(\tau)^{18}\vartheta_1(\tau,z)^2\right\rvert^2\big)\\ 
    &=-\log\big( Y^{10}\left\lvert
    p\psi(\tau,z)\big\rvert^2\right)\,,
\end{split}
\end{equation}
where we have defined 
\begin{equation}
\mathcal{I}_1^\mathrm{reg}(\Omega):=\lim_{s\rightarrow 1}\left\{
\mathcal{I}_1(\Omega,s)-
\frac{c(0)}{2}\left(\frac{1}{s-1}+\gamma_E-\log 4\pi\right)\right\}\,.
\end{equation}

Finally, if we define the regularization of \eqref{modintegral} by 
$\mathcal{I}^\mathrm{reg}(\Omega):=
\mathcal{I}_0(\Omega)+\mathcal{I}_1^\mathrm{reg}(\Omega)+
 \mathcal{I}_2(\Omega)$, it follows that
\begin{equation}
  \begin{split}
   \mathcal{I}^\mathrm{reg}(\Omega)   
  &=-\log\bigg( Y^{10}\bigg\lvert 
pqy \prod\limits_{(k,l,r)>0} (1- p^kq^ly^r)^{c(4kl-r^2)} 
\bigg\rvert^2\bigg)\\
  &=-\log\big( Y^{10}\abs{\chi_{10}(\Omega)}^2\big)\,.
  \end{split}
\end{equation}
This is the desired result.

\section{7. Symmetric products  and Igusa cusp form}
Suppose that $M$ is a smooth projective surface. Let
$M^{(m)}:=\mathrm{S}^m M$ denote the $m^\mathrm{th}$ symmetric product
of $M$'s and let $M^{[m]}:=\Hilb^m(M)$ denote the Hilbert scheme of
zero dimensional subschemes of length $m$. The Hilbert scheme
$M^{[m]}$ is a smooth resolution of $M^{(m)}$.  G{\" o}ttsche [Goe]
proved that the generating function of the Euler characteristics of
$M^{[m]}$'s is given by
\begin{equation}
  \sum_{m=0}^\infty p^m \chi(M^{[m]})= \prod_{k=1}^\infty
  \frac{1}{(1-p^k)^{\chi(M)}}\,.
\end{equation}
This, in particular, means that if $M$ is a K3 surface,
\begin{equation}
  \sum_{m=0}^\infty p^m \chi(M^{[m]})=\frac{p}{\Delta(\sigma)}\,.
\end{equation}

This formula of G{\" o}ttsche has aroused much interest [HH][VW][Nak].
In particular, Vafa and Witten [VW] gave a physical derivation of the
orbifold formula [HH] which reads for a K3 surface $M$ as
\begin{equation}
  \sum_{m=0}^\infty p^m
  \chi(M^{(m)})=\frac{p}{\Delta(\sigma)}\,.
\end{equation}
In this formula $\chi(M^{(m)})$ should be understood as the orbifold
Euler characteristic of the symmetric product $M^{(m)}$ where the
orbifold procedure is with respect to the symmetric group
$\mathfrak{S}_m$.  Vafa and Witten, based on this formula, pointed out
a mysterious connection between K3 surfaces and bosonic string.
Subsequently their observation played an important role in
understanding the duality between heterotic string compactified on
$T^4$ and type IIA string compactified on a K3 surface. See for
instance [BSV] [YZ].

In the sequel $M$ is assumed to be a K3 surface. The procedure of Vafa
and Witten for the orbifold Euler characteristic of the symmetric
products was extended for the orbifold elliptic genera of the
symmetric products by Dijkgraaf, Moore, E.~Verlinde and H.~Verlinde
[DMVV] who proved that
\begin{equation}\label{Del1}
  \sum_{m=0}^\infty p^m Z[M^{(m)}](\tau,z)=
  \prod_{\substack{k>0\\(k,l,r)>0}}\frac{1}{(1-p^kq^ly^r)^{c(4kl-r^2)}}\,,
\end{equation}
where again $Z[M^{(m)}]$ should be understood as the orbifold elliptic
genus of $M^{(m)}$.  With the use of \eqref{prtoH} this can also be
written as
\begin{equation}\label{Del2}
  \sum_{m=0}^\infty p^m Z[M^{(m)}](\tau,z)=
\exp\left(\sum_{\ell=1}^\infty p^\ell \zeta_\ell(\tau,z)\right)\,,
\end{equation}
where we have set 
\begin{equation}
  \zeta_\ell:=Z[M]\big\vert_{0,1} V_\ell\,, \quad (\ell\in \NN)\,.
\end{equation}
 Consequently, we see that
\begin{equation}
  Z[M^{(m)}]=s_m(\zeta_1,\ldots,\zeta_m)\,, \quad(m=0,1,2, \ldots)\,,
\end{equation}
where $s_m$ are the Schur polynomials.
For instance,
\begin{equation}
  \begin{split}
   &Z[M^{(1)}](\tau,z)=s_1(\zeta_1)=\zeta_1=Z(\tau,z)\,,\\
   &Z[M^{(2)}](\tau,z)=s_2(\zeta_1,\zeta_2)=\frac{1}{2}{\zeta_1}^2+\zeta_2\\
     &\hspace{1cm}=\frac{1}{2}\biggl\{Z(\tau,z)^2+Z(2\tau,2z)
+Z\left(\frac{\tau}{2},z\right)+Z\left(\frac{\tau+1}{2},z\right)\biggr\}\,,
  \end{split}
\end{equation}
where $Z(\tau,z)=Z[M](\tau,z)$. In general 
$Z[M^{(m)}](\tau,z)$ contains the term $\frac{1}{m!}Z(\tau,z)^m$ as
the contribution from the untwisted sector.

The Hilbert scheme $M^{[m]}$ of a K3 surface $M$ is an $2m$
dimensional Calabi-Yau manifold [Bea].  Thus it is tempting to think
that one can replace $Z[M^{(m)}]$ by $Z[M^{[m]}]$ in the above
equations. This is true at least at the level of
$\chi_y$-genera\footnote{On the contrary to the misleading impression
  given in [KYY] and [KM], the information of the $\chi_y$-genus is
  {\it not\/} enough to determine the elliptic genus completely if the
  index of the elliptic genus as a weak Jacobi form, {\it i.e.\/} the
  dimension of the manifold or the rank of the bundle is sufficiently
  large. This can be seen for instance from Corollary \ref{stth}.
  Since, for a $2m$ dimensional Calabi-Yau manifold, the elliptic
  genus $Z(\tau,z)$ is a weak Jacobi form of weight $0$ and index $m$,
  it must be of the form $p_{2m}(E_4,E_6,\wp)K^{2m}$ according to
  Corollary \ref{stth}.  However, the $\chi_y$-genus in general does
  not fix the polynomial $p_{2m}$ uniquely if $m$ is large enough.  I
  am grateful to V. Gritsenko for discussion on this point.}.  To see
this, we take the limit $q\rightarrow 0$ in \eqref{Del1} or
\eqref{Del2} to obtain that
\begin{equation}
  \begin{split}
    \sum_{m=0}^\infty (p/y)^m \chi_y(M^{(m)}) &=\prod_{k=1}^\infty
    \frac{1}{(1-p^k)^{20}(1-p^ky)^2(1-p^ky^{-1})^2}\\[1mm]
    &=\exp\left(\sum_{a=1}^\infty
      \frac{1}{a}(p/y)^a\frac{\chi_{y^a}(M)}{1-p^a}\right)\,.
\end{split}
\end{equation}
That this result with the replacement of $\chi_y(M^{(m)})$ by
$\chi_y(M^{[m]})$ holds was proved in [GS][Che]. 

String theorists are very familiar to the Igusa cusp form $\chi_{10}$
since it appears in the two-loop vacuum amplitude of bosonic string
[Two].  Now I wish to provide an interpretation of \eqref{Del1} from
this viewpoint.  Eq.\eqref{Del1} can be more suggestively rewritten as
\begin{equation}\label{pinch}
  \frac{1}{\chi_{10}(\Omega)}=\sum_{m=0}^\infty p^{m-1}
  \frac{Z[M^{(m)}](\tau,z)K(\tau,z)^{-2}}{\Delta(\tau)}\,,
\end{equation}
where we used \eqref{borcherds}, \eqref{psiexp} and \eqref{Del2}.  The
LHS is the chiral part of the integrand appearing in the calculation
of the two-loop vacuum amplitude of bosonic string.  Since being
expanded in powers of $p$, the RHS should represent the contributions
of the one-loop two-point functions of the physical states at various
mass levels according to the factorization principle when one of the
two handles of the genus two Riemann surface is very thin and long.

As is well-known the space of physical states of the critical (open)
bosonic string can be identified with the BRST cohomology group of
ghost number $1$ which is graded by level $m\in \NN_0$ and  momentum
$k_m\in \RR^{25,1}$  satisfying the mass shell
conditions
\begin{equation}
  \frac{(k_m,k_m)}{2}+m=1\,.
\end{equation}
 Let $e_1,\ldots,e_{26}$ be the basis of $
\RR^{25,1}$ whose Gramm matrix is given by
$\kappa:=\mathrm{diag}(+1,\ldots,+1,-1)$. The components of vectors in
$\RR^{25,1}$ with respect to this basis are represented by Greek
superscripts and lowered by the metric $\kappa$ as usual. 
 Let $c(z)$ be the ghost field and let $X(z)$ denote chiral free
bosons taking values in $\RR^{25,1}$.  The representatives of the BRST
cohomology classes are of the form $c(z){\cal V}^m[v^\alpha_*,k_m](z)$
$(\alpha=1,\ldots,p_{24}(m))$ where $q/\Delta(\tau)=\sum_{m=0}^\infty
p_{24}(m)q^m$ and 
\begin{equation}\label{vertex}
 {\cal V}^m[v^\alpha_*,k_m](z)= 
:U_m(v^\alpha_*,X)(z)e^{\sqrt{-1}(k_m,X(z))}:\,,\quad 
\alpha=1,2,\ldots,p_{24}(m)\,,
\end{equation}
with $U_0(v^\alpha_*,X)(z)=1$ and 
\begin{equation}
  U_m(v^\alpha_*,X)(z):=\sum_{n=1}^m \sum_{\mu_1,\ldots,\mu_n}\,
  \sum_{\substack{m_1\ge 1,\ldots,m_n\ge 1\\[.5mm] m_1+\cdots+m_n=m}}
  v^{\alpha}_{\mu_1,\ldots,\mu_n}\prod_{j=1}^n\sqrt{-1}\,
  \partial^{m_j}X^{\mu_j}(z)\,,
\end{equation}
for $m\ge 1$.  The polarization tensors
$v^{\alpha}_{\mu_1,\ldots,\mu_n}$ are such that ${\cal
  V}^m[v^\alpha_*,k_m](z)$ is primary and of conformal weight
$1$. There is a degree of freedom for the choice of the set of
polarization tensors corresponding to that for the choice of a
representative in the BRST cohomology class.  The vertex operators
${\cal V}^m[v^\alpha_*,k_m](z)$ correspond to the emissions of
 particles whose $(\mathrm{mass})^2$ are equal to $2m-2$\footnote{Here
  we are working in the convention of open string theory.}.

Thus if we believe the factorization principle, the expansion
\eqref{pinch} suggests 
\begin{multline}\label{factorization}
  Z[M^{(m)}](\tau,z)K(\tau,z)^{-2}=\\
\frac{1}{(2\pi\sqrt{-1})^2}
  \sum_{\alpha=1}^{p_{24}(m)} \left< {\cal V}^m[v^\alpha_*,k_m](z) \,
    {\cal V}^m[v^\alpha_*,-k_m](0)\right>\,,
\end{multline}
where the brackets stand for correlation functions on the elliptic
curve parametrized by $\tau$.  This expression must be viewed with
some care. In the computations of string amplitudes we can use
arbitrary representatives of the BRST cohomology classes since the
differences can be expressed as total derivatives in the moduli space
of the worldsheet Riemann surface. However here we are dealing with
the (chiral part of) integrand and not the integral over the moduli
space. Thus the polarization tensors cannot be arbitrary even in the
same BRST cohomology classes and must correspond to {\it special\/}
choices of representatives associated with the expansion
\eqref{pinch}.

Although we do not know these polarization tensors precisely  we can
make some general statements about the evaluation of the RHS of
\eqref{factorization}.
First notice that the function
$\hat K(\tau,z):=K(\tau,z)/(2\pi \sqrt{-1})$, where $K(\tau,z)$ is as
defined in \eqref{Kdef}, is the {\it prime form\/} [Mum] behaving $\hat
K(\tau,z)\sim z$ as $z\rightarrow 0$. Therefore we have
\begin{equation}
  \left<:e^{\sqrt{-1}(k_m,X(z))}::e^{-\sqrt{-1}(k_m,X(0))}:\right>
=\hat K(\tau,z)^{-(k_m,k_m)}=\hat K(\tau,z)^{2m-2}\,.
\end{equation}
Since the elliptic genera $Z[M^{(m)}]$ (as well as $Z[M^{[m]}]$) are
weak Jacobi forms of weight $0$ and index $m$ and hence they must be
of the form $p_{2m}(E_4,E_6,\wp)K^{2m}$ according to Corollary
\ref{stth}. Thus the weight $2m$ homogeneous polynomial
$p_{2m}(E_4,E_6,\wp)$ should stem from the contractions other than
that between $e^{\sqrt{-1}(k_m,X)}(z)$ and $e^{-\sqrt{-1}(k_m,X)}(0)$
in the evaluation of $ \left< {\cal V}^m[v^\alpha_*,k_m](z) \, {\cal
    V}^m[v^\alpha_*,-k_m](0)\right>/(2\pi \sqrt{-1})^2$.

We recall  that 
$\wp$-function satisfies  the differential equations
\begin{align}
  {D_y}^2\wp&=6\, \wp^2-\frac{1}{24}\, E_4\,,\\ {D_y}^3\wp&=12\, \wp\cdot
  D_y\wp\,,
\end{align}
in addition to \eqref{perel}. Thus any weight $2m$ homogeneous
polynomial of $E_4$, $E_6$ and $\wp$ can be expressed as a linear
combination of
\begin{equation}\label{peproduct}
  \prod_{i\ge 0}({D_y}^{i}\wp)^{a_i}\,,
\end{equation}
where $a_i$ are nonnegative integers such that $\sum_{i\ge
  0}(i+2)a_i=2m$ and vice versa. The appearance of the terms of the
form \eqref{peproduct} can be understood at least for the contraction
between $U_m(v^\alpha_*,X)(z)$ and $U_m(v^\alpha_*,X)(0)$ if we adopt 
the rule
\begin{equation}
  \left<\partial X^\mu(z)\partial X^\nu(0)\right>
  \longrightarrow - \kappa^{\mu\nu}\hat \wp(\tau,z)\,,
\end{equation}
where $\hat \wp(\tau,z):=(2\pi \sqrt{-1})^2 \wp(\tau,z)$ and we remind
that the $\wp$-function satisfies
\begin{equation}
  \wp(\tau,z)=-{D_y}^2\log K(\tau,z)+
\frac{1}{12}E_2(\tau)=\frac{1}{(\log y)^2}+\cdots\,.
\end{equation}

As an example, consider the massless case $(m=1)$. There are
$p_{24}(1)=24$ polarization (contravariant) vectors $v^\alpha$
satisfying $(v^\alpha,v^\beta)=\delta^{\alpha\beta}$. They are
orthogonal to the momentum $k_1$. Thus we see that
\begin{equation}
  \begin{split}
    -\sum_{\alpha=1}^{24} \sum_{\mu,\nu=1}^{26} \left<v^\alpha_\mu
      \partial X^\mu(z):e^{\sqrt{-1}(k_1,X(z))}:v^\alpha_\nu \partial
      X^\nu(z):e^{-\sqrt{-1}(k_1,X(0))}:\right>\\=24\hat
    \wp(\tau,z)\hat K(\tau,z)^{-(k_1,k_1)}=24\hat \wp(\tau,z)\,.
\end{split}
\end{equation}
Hence we have 
\begin{equation}
  Z[M^{(1)}](\tau,z)=\frac{1}{(2\pi \sqrt{-1})^2}\cdot
  24\hat\wp(\tau,z)K(\tau,z)^2 =24\wp(\tau,z)\,K(\tau,z)^{2}\,.
\end{equation}
However this is precisely the expression \eqref{massless} we
encountered before!

I have calculated several cases explicitly to find  the results:
\smallskip
\begin{center}
\begin{tabular}{c|l}
$m$&$Z[M^{(m)}](\tau,z)$\\ \hline
$1$&$24\, \wp\, K^2$\\[1mm]
$2$&$ (324\, \wp^2+\frac{3}{4}\, E_4)\, K^4$\\[1mm]
$3$&$ (3200\, \wp^3+\frac{64}{3}\,  E_4\,\wp +\frac{10}{27}\, E_6)\,
 K^6$\\[1mm]
$4$&$ (25650\, \wp^4+\frac{1329}{4}\, E_4\,\wp^2+\frac{45}{4}\, E_6\, \wp
+\frac{199}{384}\, {E_4}^2)\, K^8$\\[1mm]
$5$&$ (176256\, \wp^5+3720\, E_4\, \wp^3+186\, E_6\,\wp^2
+17\, {E_4}^2\,\wp+\frac{31}{72}\, E_4\, E_6) \,K^{10}$ \\ 
  \end{tabular}
\end{center}

{\noindent For} these lower values of $m$'s the information of
$\chi_y$-genera was sufficient to determine the elliptic genera and
hence we can without doubt replace $M^{(m)}$ by $M^{[m]}$ in the
table. Notice that the coefficient of $\wp^{m}K^{2m}$ coincides with
$p_{24}(m)$.  This must be so since $\wp^{m}K^{2m}$ tends to $1$ as
$z\rightarrow 0$ while the rest of terms vanish in this limit.

To summarize the situation, {\it the elliptic genus $Z[M^{(m)}]$ is
  equal to the sum of the (chiral) two-point functions of all the
  physical states having the mass squared $2m-2$ on an elliptic curve,
  divided by the (chiral) tachyon two-point function on the same
  elliptic curve}.

It is yet to be seen how much this statement is of value for a further
understanding of string duality.

Finally we recall that in addition to the relation between bosonic
string and K3 surfaces Vafa and Witten [VW] pointed out the one
between superstring and abelian surfaces. The vacuum two-loop as well
as one-loop amplitude of superstring vanishes and there is no analog
of $\chi_{10}$. This is in parallel with the fact that the elliptic
genus of abelian surfaces vanishes identically.

\section{8. Discussion}
Perhaps one of the still open problems regarding the phenomena we
observed in this paper is to clarify what kind of algebras are behind
them and what kind of roles these algebras are playing. Gritsenko and
Nikulin associated with $\chi_{10}$ or more precisely $\Delta_5$, a
generalized Kac-Moody superalgebra whose real simple roots are
characterized by the Cartan matrix
\begin{equation}
  \begin{pmatrix}
    2&-2&-2\\
    -2&2&-2\\
    -2&-2&2
  \end{pmatrix}\,.
\end{equation}
As a submatrix, this Cartan matrix contains (in two possible ways)
that of $A_1^{(1)}$
\begin{equation}
  \begin{pmatrix}
    2&-2\\
-2&2
  \end{pmatrix}\,.
\end{equation}
This affine Lie algebra $A_1^{(1)}$ has a very clear physical
origin. If $M$ is a K3 surface $M^{[m]}$ is a hyperk{\" a}hler
manifold [Bea] of complex dimension $2m$. The elliptic genus
$Z[M^{[m]}]$ is associated with the sigma model with $M^{[m]}$ as the
target space. This sigma model is governed by the $N=4$ superconformal
algebra which has a Virasoro central charge $6m$ and contains
$A_1^{(1)}$ at level $m$ as a subalgebra. This is the physical origin
of why the elliptic genus $Z[M^{[m]}]$ as a weak Jacobi form of index
$m$ can be expanded in terms of the $A_1^{(1)}$ theta functions as in
\eqref{expA1}. Thus the generalized Kac-Moody superalgebra contains a
tower of integrable representations of $A_1^{(1)}$ at arbitrary
levels.  This is reminiscent of the situation in [FF].

On the other hand, if we take the viewpoint in the last section
seriously, $N=4$ superconformal invariance must somehow be realized at
each mass level of bosonic string theory. This expectation clearly
needs further investigation.

As is well-known, the tachyon vertex operators of bosonic string are
utilized in the Frenkel-Kac-Segal constructions for the basic
representations of (simply-laced) affine Lie algebras.  It has long
been anticipated by many  that the massless as well as massive
vertex operators might also play roles in realizing some larger
algebraic structures.  Although the situation discussed in this paper
may not be the right one for this anticipation since, for instance, we
were dealing with the {\it uncompactified\/} bosonic string, it is
still interesting to observe that with a genus two Riemann surface is
associated the big algebraic structure of the Gritsenko-Nikulin
algebra and if we pinch one of the handle of the Riemann surface and
use the factorization principle we see that vertex operators of all
mass levels appear naturally.

\pagestyle{myheadings} \markboth{{}\myaddressfont \hfil T. Kawai}
{\myaddressfont K3 surfaces and Igusa cusp form\hfil {} }
 
\vskip 1pc \noindent{\it Acknowledgments.} I should like to express my
gratitude to the Taniguchi foundation and the organizers of this
symposium. I especially thank K. Saito for discussion at various
occasions.  I am also grateful to  V. Gritsenko for clarifying
correspondence and to Y. Yamada for helpful suggestion.

\def\thebibliography#1{\vskip 1.2pc{\centerline {\bf References}}\vskip 4pt
\list
 {[\arabic{enumi}]}{\settowidth\labelwidth{[#1]}\leftmargin\labelwidth
 \advance\leftmargin\labelsep
 \usecounter{enumi}}
 \def\newblock{\hskip .11em plus .33em minus .07em}
 \sloppy\clubpenalty4000\widowpenalty4000
 \sfcode`\.=1000\relax}
\let\endthebibliography=\endlist

%\begin{thebibliography}{ABCD}

\vskip 2pc
%\pagebreak[4]
\begin{flushleft}
Toshiya Kawai \nopagebreak \\ 
Research Institute for Mathematical Sciences \nopagebreak \\ 
Kyoto University \nopagebreak\\ 
Kyoto 606-01, Japan \nopagebreak\\ 
e-mail: {\tt toshiya@kurims.kyoto-u.ac.jp}  
\end{flushleft}
\vskip 6pt
%\noindent Received 


\begin{thebibliography}{ }

\bibitem[Apo]{Apo} T.M. Apostol, {\it Modular Functions and Dirichlet
    Series in Number Theory}, Graduate Texts in Mathematics {\bf 41},
  Springer-Verlag, 1976.

\bibitem[Bea]{Bea} A. Beauville, {\it Vari{\' e}t{\' e}s k{\"
      a}hleriennes dont la premi{\` e}re classe de Chern est nulle},
  J. Diff. Geom. {\bf 18} (1983), 755--782.

\bibitem[Bor1]{Bor1} R.E. Borcherds, {\it Automorphic forms on ${\rm
      O}\sb {s+2,2}({\bf R})$ and infinite products},
  Invent. Math. {\bf 120} (1995), 161--213.

\bibitem[Bor2]{Bor2} R.E. Borcherds, {\it Automorphic forms with
    singularities on Grassmannians}, alg-geom/9609022.

\bibitem[BSV]{BSV} M. Bershadsky, V. Sadov and C. Vafa, {\it D-Branes
    and Topological Field Theories}, Nucl. Phys. {\bf B463} (1996)
  420--434.

\bibitem[CCL]{CCL} G.L. Cardoso, G. Curio and D. Lust, {\it
    Perturbative couplings and modular forms in N=2 string models with
    a Wilson line}, Nucl. Phys. B491 (1997) 147-183.

\bibitem[Che]{Che} J. Cheah, {\it On the cohomology of Hilbert schemes of
    points}, J. Alg. Geom. {\bf 5} (1996), 479--511.

\bibitem[DKL]{DKL} L.J. Dixon, V.  Kaplunovsky and J.  Louis, {\it
    Moduli dependence of string loop corrections to gauge coupling
    constants}, Nucl. Phys. {\bf B355} (1991) 649--688.

\bibitem[DMVV]{DMVV} R. Dijkgraaf, G. Moore, E. Verlinde and
  H. Verlinde, {\it Elliptic Genera of Symmetric Products and Second
    Quantized Strings}, Commun. Math. Phys. {\bf 185} (1997) 197--209.


\bibitem[DVV]{DVV} R. Dijkgraaf, E. Verlinde and H. Verlinde, {\it
    Counting Dyons in N=4 String Theory}, Nucl. Phys. {\bf B484}
  (1997) 543-561.

\bibitem[EMOT]{EMOT} A. Erd{\' e}li, W. Magnus, F. Oberhettinger and
  F.G. Tricomi, {\it Higher Transcendental Functions}, Vol. 2,
  McGraw-Hill, 1953.

\bibitem[EOTY]{EOTY} T. Eguchi, H. Ooguri, A. Taormina and S.-K. Yang,
  {\it Superconformal Algebras and String Compactification on
    Manifolds with SU(n) Holonomy}, Nucl.~Phys. {\bf B315} (1989) 193--221.


\bibitem[EZ]{EZ} M. Eichler and D. Zagier, {\it The Theory of Jacobi
    Forms}, Progress in Mathematics {\bf 55}, Birkh\"{a}user, 1985.

\bibitem[FF]{FF} A.J. Feingold and I.B. Frenkel, {\it A hyperbolic
    Kac-Moody algebra and the theory of Siegel modular forms of genus
    2}, Math. Ann. {\bf 263} (1983) 87--114.

\bibitem[Fre]{Fre} E. Freitag, {\it Siegelsche Modulformen},
  Grundlehren der mathematischen Wissenschaften, Bd. {\bf 254},
  Springer-Verlag, 1983;\ {\it Singular Modular Forms and Theta
    Relations}, Lecture Notes in Mathematics {\bf 1487},
  Springer-Verlag, 1991.

\bibitem[GN1]{GN1} V.A.~Gritsenko and V.V.~Nikulin, {\it Siegel
    automorphic form corrections of some Lorentzian Kac--Moody Lie
    algebras}, Amer. J. Math. {\bf 119} (1997), 181--224, alg-geom/9504006;
  {\it The Igusa modular forms and ``the simplest'' Lorentzian
    Kac--Moody algebras}, alg-geom/9603010.

\bibitem[GN2]{GN2} V.A.~Gritsenko and V.V.~Nikulin, {\it Automorphic
    Forms and Lorentzian Kac--Moody Algebras}, part I and part II,
  alg-geom/9610022 and alg-geom/9611028.

\bibitem[Goe]{Goe} L. G{\" o}ttsche, {\it The Betti numbers of the
    Hilbert Schemes of Points on a Smooth Projective Surface},
  Math. Ann. {\bf 286} (1990) 193--207;\ {\it Hilbert Schemes of
    Zero-dimensional Subschemes of Smooth Varieties}, Lecture Notes in
  Mathematics {\bf 1572}, Springer-Verlag, 1994.

\bibitem[GS]{GS} L. G{\" o}ttsche and W. Soergel, {\it Perverse
    sheaves and the cohomology of Hilbert schemes of smooth algebraic
    surfaces}, Math. Ann. {\bf 296} (1993), 235--245.

\bibitem[HH]{HH} F. Hirzebruch and T. H{\" o}ffer, {\it On the Euler
    number of an Orbifold}, Math. Ann. {\bf 286} (1990) 255--266.

\bibitem[HM]{HM} J.A. Harvey and G. Moore, {\it Algebras, BPS States,
    and Strings}, Nucl. Phys. {\bf B463} (1996) 315--368.

\bibitem[Igu]{Igu} J. Igusa, {\it On Siegel Modular Forms of Genus
    Two}, Amer. J. Math. {\bf 84} (1962) 175--200;\ {\it On Siegel
    Modular Forms of Genus Two (II)}, Amer. J. Math. {\bf 86} (1964)
  392--412.

\bibitem[Kaw1]{Kaw1} T. Kawai, {\it $N=2$ heterotic string threshold
    correction, $K3$ surface and generalized Kac-Moody superalgebra},
  Phys. Lett. {\bf B372} (1996) 59--64.


\bibitem[Kaw2]{Kaw2} T. Kawai, {\it String duality and modular forms},
Phys. Lett. {\bf  B397} (1997) 51--62.

\bibitem[KM]{KM} T. Kawai and K. Mohri, {\it Geometry of (0,2)
    Landau-Ginzburg Orbifolds}, Nucl. Phys. {\bf B425} (1994) 191--216.

\bibitem[KYY]{KYY} T. Kawai, Y. Yamada and S.-K. Yang, {\it Elliptic
    Genera and N=2 Superconformal Field Theory}, Nucl. Phys. {\bf
    B414} (1994) 191--212.

\bibitem[Lan1]{Lan1} P.S. Landweber (Ed.), {\it Elliptic Curves and
    Modular Forms in Algebraic Topology}, Lecture Notes in
  Mathematics, {\bf 1326}, Springer-Verlag, 1988.

\bibitem[Lan2]{Lan2} S. Lang, {\it Elliptic Functions}, Addison-Wesley,
  1973.  


\bibitem[MS]{MS} P. Mayr and S. Stieberger, {\it Moduli dependence of
    one-loop gauge couplings in (0,2) compactifications},
  Phys. Lett. B355 (1995) 107--116.

\bibitem[Mum]{Mum} D. Mumford, {\it Tata Lectures on Theta II},
  Birkh{\" a}user, 1984.

\bibitem[Nak]{Nak} H. Nakajima, {\it Heisenberg algebra and
    Hilbert schemes of points on projective surfaces},
  alg-geom/9507012.

\bibitem[Sie]{Sie} C.L. Siegel, {\it Advanced Analytic Number Theory},
  Studies in Mathematics {\bf 9}, Tata Institute of Fundamental
  Research, 1980.

\bibitem[Two]{Two} A.A. Belavin and V.G. Knizhnik, {\it Complex
    geometry and the theory of quantum strings}, Sov. Phys. JETP {\bf
    64} (1986) 214--228, Zh. Eksp. Teor. Fiz. {\bf 191} (1986)
  364--390;\ {\it Algebraic geometry and the geometry of quantum
    strings}, Phys. Lett. {\bf 168B} 201--206.\\
 A. Belavin,
  V. Knizhnik, A. Morozov and A. Perelomov, {\it Two and three loop
    amplitudes in the bosonic string theory}, JETP Lett. {\bf 43}
  (1986) 411--414, Phys. Lett. {\bf B177} (1986) 324--328.\\
 G. Moore, {\it
    Modular forms and two loop string physics}, Phys. Lett. {\bf B176}
  (1986) 369--379.\\
 A. Kato, Y. Matsuo and S. Odake, {\it Modular
    invariance and two loop bosonic string vacuum amplitude},
  Phys. Lett. {\bf B179} (1986) 241--246.

\bibitem[VW]{VW} C. Vafa and E. Witten, {\it A strong coupling test of
    S-duality}, Nucl. Phys. {\bf 431} (1994) 3--77.

\bibitem[Wal]{Wal} M.A.  Walton, {\it The Heterotic String on the
    Simplest Calabi-Yau Manifold and Its Orbifold Limits},
  Phys.~Rev. {\bf D37} (1988) 377--390.

\bibitem[Wit]{Wit} E. Witten, {\it The index of the Dirac Operator in
    Loop Space}, pp. 161--181 in [Lan2].

\bibitem[YZ]{YZ} S-T. Yau and E. Zaslow, {\it BPS States, String
    Duality, and Nodal Curves on K3}, Nucl. Phys. {\bf B471} (1996)
  503--512.
\end{thebibliography}
\end{document}